
\hsize=6.5in
\vsize=9in
\def\fun#1#2{\lower3.6pt\vbox{\baselineskip0pt\lineskip.9pt
  \ialign{$\mathsurround=0pt#1\hfil##\hfil$\crcr#2\crcr\sim\crcr}}}
\def\la{\mathrel{\mathpalette\fun <}}
\def\ga{\mathrel{\mathpalette\fun >}}
\baselineskip 7truemm plus1truemm
\xspaceskip 8pt plus 3pt minus 3pt
\parskip=3truemm
\space\
\vskip 0.5in
\centerline{\bf CONSTRAINTS ON NEUTRINO OSCILLATIONS}
\centerline{\bf  FROM BIG BANG NUCLEOSYNETHESIS}
\vskip .7cm
\centerline{\bf X. Shi$^1$, D. N. Schramm$^{1, 2}$, B. D. Fields$^1$}
\vskip 0.3cm
\centerline{\sl $^1$The University of Chicago,}
\centerline{\sl Chicago, IL 60637-1433}
\vskip .3cm
\centerline{\sl $^2$NASA/Fermilab Astrophysics Center,}
\centerline{\sl Fermi National Accelerator Laboratory,}
\centerline{\sl Box 500, Batavia, IL 60510-500}
\vskip 0.2in
\centerline{ABSTRACT}
\vskip 0.2in
\noindent We discuss in detail the effect of neutrino oscillations
in Big Bang nucleosynthesis, between active and sterile neutrinos,
as well as between active and active neutrinos.
We calculate the constraints on mixings between active and sterile
neutrinos from the present observation of the primordial helium abundance
and discuss the potential implications
on various astrophysical and cosmological
problems of such oscillations. In  particular, we show that
large angle sterile neutrino mixing seems to be excluded
as a MSW solution to the solar neutrino situation or a solution to
the atmospheric neutrino mixing hinted at in some underground experiments.
We show how with this constraint, the next generation of solar neutrino
experiments should be able to determine the resolution of the solar neutrino
problem. It is also shown how sterile neutrinos remain a viable dark matter
candidate.
\vskip .8in
\centerline{PACS numbers: 12.15.Ff, 98.80.Ft, 96.60.Kx, 95.35.+d}

\vfill\eject
\baselineskip=20pt
\centerline{\bf I. Introduction}
Neutrino mixing and oscillations have long been one of the most important
and exciting topics in physics since they directly involve
physics beyond the standard model [1].
While direct evidence of oscillations has yet to be found,
there is some indirect experimental evidence.
Perhaps most intriguing are
the solar neutrino experiments which observed a solar neutrino
flux that is significantly lower than the prediction of the standard
solar models [2]. In particular, the result of the Homestake experiment is a
factor of 3 and more than 5$\sigma$ below the predictions
of the standard solar model of Bahcall {\sl et al.} [3], and
also significantly
lower than the standard solar model of Turck-Chi\`eze {\sl et al.} [4].
This observed deficit in fluxes, if real, can be readily
explained by neutrino mixing and oscillation [5,6].
Another possible hint of neutrino oscillation (in a different parameter
range) comes from the atmospheric neutrino
experiments at Kamioka [7] and IMB [8] and recently at Soudan II [9],
which observed deficits in the atmospheric $\nu_\mu$ flux relative
to the $\nu_e$ flux.
Both of these experimental indications are subject to controversy,
especially the atmospheric neutrino experiments.
However, it is nonetheless useful to explore
the effects of neutrino oscillations in various astrophysical and cosmological
environments.

There are two basic neutrino oscillation solutions to the solar neutrino
problem. One is the vaccum mixing (``just-so mixing''),
in which solar neutrinos $\nu_e$ simply oscillate into either active
($\nu_\mu$ and $\nu_\tau$) or sterile neutrinos ($\nu_s$)
on their way to the earth [10]. It requires a relatively large mixing angle
$\theta$ ($sin^22\theta\ga 0.75$) and a mass square difference
between the two species of about 10$^{-10}$ eV$^2$.
The other solution which is much more robust in that it allows
a larger set of parameter space, is the
MSW matter mixing solution, in which $\nu_e$'s
go through a resonance inside the
Sun and change into other neutrino species (again either active or sterile)
[11].
Two possible ranges of mixing parameters exist
in this mechanism to simultaneously explain the observed neutrino
deficit in each of the operating experiments [12]:
the small mixing angle solution with $\delta m^2=m_2^2-m_1^2\sim
10^{-6}$ to $10^{-5}$ eV$^2$ and $sin^22\theta\sim 5\times 10^{-3}$,
and the large mixing angle solution with
$\delta m^2=m_2^2-m_1^2\sim 10^{-6}$ to $10^{-4}$ eV$^2$
and $sin^22\theta\ga 0.4$. On the other hand,
To solve the reported atmospheric $\nu_\mu$ deficit, a vaccum mixing
between $\nu_\mu$ and either $\nu_\tau$ or $\nu_s$ is needed,
with a large mixing $sin^22\theta\ga 0.4$ and $\delta m^2$ of order $10^{-3}$
to $10^{-1}$ eV$^2$ [13].

Various terrestrial experiments searching for neutrino oscillations
have set stringent bounds on the mixing parameters [14]. However, they
were unable to explore the mixing parameter space
with either small mass square difference (with $\delta m^2<10^{-2}$ eV$^2$
in $\nu_e$-$\nu_x$ mixing, and $\delta m^2<10^{-1}$ eV$^2$ in $\nu_\mu$
-$\nu_x$ mixing),
or small mixing angles with $sin^2\theta\la 10^{-2}$ to $10^{-1}$,
where $\theta$ is the mixing angle.
Cosmological and astrophysical considerations, however,
give more stringent constraints on these parts of the mixing parameter space.
One such constraint comes from consideration of Big Bang nucleosynthesis
(hereafter BBN).

The effect of neutrino mixing between a sterile neutrino
and an active neutrino on BBN
has been investigated by various groups. The numerical calculation of
Enqvist {\sl et al.} [15] has excluded a large area on the mixing parameter
space, including the $\nu_\mu$-$\nu_s$ mixing solution to the atmospheric
neutrino problem and a large portion of the large angle $\nu_e$-$\nu_s$
MSW mixing solution to the solar neutrino problem.
Analytical calculations exploring the question were done by
Barbieri {\sl et al.} [16] and Cline [17].
The possible effect of $\nu_e$-$\nu_{\mu} ({\rm or}\ \nu_\tau)$ mixing
on BBN has also been estimated by Langacker
{\sl et al.}, with a conclusion that the effect might be too small to
set constraints [18]. Since the number densities of $\nu_e$ and $\nu_\mu$
(or $\nu_\tau$) are nearly equal in the primordial nucleosynthesis
epoch, it is anticipated that a mixing between $\nu_e$ and $\nu_\mu$ (or
$\nu_\tau$) would not change the predictions of standard
BBN to a degree that is currently observable.

In this paper we attempt to discuss the neutrino mixing in further
detail and calculate the constraints on the mixing parameters using
a full numerical BBN study with the
latest primordial helium abundance observations. In particular the
new $^4$He abundance determinations have increased with
respect to the previously
used numbers to an extent that could alter the previous conclusions.
We will discuss the implications of the allowed
neutrino oscillations on the solar neutrino
problem, the atmospheric problem and the dark matter problem of the universe.

\centerline{\bf II. Neutrino Oscillation Formalism}

In the case of a two-family neutrino oscillation between $\nu_\alpha$ and
$\nu_\beta$, we assume the mass eigenvalues of the two
mass eigenstates $\nu_1$ and $\nu_2$ to be $m_1$ and $m_2$, and
the mixing angle to be $\theta$. Then the time evolution of a neutrino state
is discribed by eq. (1) [19]:
$$i{d\over dt}\left(\matrix{C_\alpha(t)\cr
C_\beta(t)\cr}\right)=M\left(\matrix{C_\alpha(t)\cr C_\beta(t)\cr}\right),\eqno
  (1)$$
where in the vacuum mixing,
$$M={1\over 4p}\left(\matrix{-\delta m^2 cos2\theta&\delta m^2
sin2\theta\cr \delta m^2 sin2\theta&\delta m^2 cos2\theta\cr}
\right).\eqno (2)$$
$p$ is the momentum of the neutrino and is assumed to be much larger than
$m_1$ and $m_2$ (relativistic regime). When there is MSW matter mixing,
$$\eqalignno{M&={1\over 4p}
\left(\matrix{-\delta m^2 cos2\theta+2p(V_\alpha-V_\beta)&\delta m^2
sin2\theta\cr \delta m^2 sin2\theta&\delta m^2 cos2\theta\cr}\right)\cr
&={\delta m_M^2\over 4p}
\left(\matrix{-cos2\theta_M&sin2\theta_M\cr sin2\theta_M&cos2\theta_M
\cr}\right)&(3)\cr}$$
where $V_\alpha$ and $V_\beta$ are the effective potential due to the
MSW effect [11,20].
$$\delta m_M^2=\sqrt{(\delta m^2sin2\theta)^2+[\delta
m^2cos2\theta+2p(V_\alpha-V_\beta)]^2}\eqno (4)$$
and $\theta_M$, the matter mixing angle, is given by
$$\theta_M={1\over 2}tan^{-1}\biggl\lbrace{tan2\theta
\over 1-[2p(V_\alpha-V_\beta)/\delta m^2]sec2\theta}\biggr\rbrace
.\eqno (5)$$

The effective potentials are obtained by calculating the
corrections to the neutrino self-energy from the ambient matter [20].
In the stellar interior (e.g., the sun),
the effective potentials can be calculated using
zero temperature field theory [20]:
$$V_e=\sqrt{2}G_F(N_e-0.5N_n),\quad
V_\mu=V_\tau=\sqrt{2}G_F(-0.5N_n),\eqno (6)$$
where $N_e$ and $N_n$ are the number density of electrons and neutrons.
In the early universe after $T<100$ MeV when $\mu$ and $\tau$ are absent,
and in the case that neutrinos are in kinetic equilibrium but may not be
in chemical equilibrium, finite temperature field theory gives [20]
$$V_e=\sqrt{2}G_FN_\gamma\bigl[L^{(e)}-12.61{pT_\gamma\over M_W^2}
-12.61{pT_{\nu_e}^4\over 4M_Z^2T_\gamma^3}[n_{\nu_e}+n_{\bar\nu_e}])\bigr],
\eqno (7)$$
$$V_\mu=\sqrt{2}G_FN_\gamma\bigl[L^{(\mu)}-12.61{pT_{\nu_\mu}^4\over 4M_Z^2
T_\gamma^3}(n_{\nu_\mu}+n_{\bar\nu_\mu})\bigr],\eqno (8)$$
where
$$L^{(e)}=({1\over 2}+2x_W)L_e+({1\over 2}-2x_W)L_p-{1\over 2}L_n+
2L_{\nu_e}+L_{\nu_\mu}+L_{\nu_\tau},\eqno (9)$$
$$L^{(\mu)}=L^{(e)}-L_e-L_{\nu_e}+L_{\nu_\mu},\eqno (10)$$
and
$$L_{\alpha}={{N_\alpha-N_{\bar\alpha}}\over N_\gamma},\eqno (11)$$
where $x_W=sin^2\theta_W=0.23$. $N_\alpha$ is the number density
of the species $\alpha$. $n_\nu=7N_\nu/(8N_\gamma)$.
A similar expression exists for $V_\tau$
if one substitutes $\mu$ to $\tau$ in eq. (8) and (10). For anti-neutrinos, one
simply substitutes $L^{(\alpha)}$ by $-L^{(\alpha)}$ to get their
effective potentials. If $L^{(a)}$ is at the level of $10^{-10}$ as
$\eta$ (=$N_B/N_\gamma=(N_n+N_p)/N_\gamma$) is,
the terms involving $L^{(a)}$ can always be neglected as long
as $T_\gamma\sim T_\nu>0.1$ MeV, which is the range in which
we are interested. In all cases, since $\nu_s$ is sterile, $V_s\equiv 0.$

When the neutrino ensemble is incoherent, the
phase of the neutrino states are not of concern.
Therefore an oscillation
is characterized by the oscillation length
$$L_{osc}={4\pi p\over\delta m_M^2}\eqno (12)$$
and the mixing angle $\theta_M$ (where $\delta m_M^2$ and $\theta_M$ reduce to
$\delta m^2$ and $\theta$ in vaccum oscillation).
If one denotes $P_{\alpha\alpha}$ as the probability of finding a $\nu_\alpha$
in a state that is initially $\nu_\alpha$,
in vaccum (or a constant medium where $V_\alpha$ and $V_\beta$ are constant),
after averaging over phases,
$$P_{\alpha\alpha}=1-{1\over 2}sin^22\theta (or\ 2\theta_M)\quad {\rm
for\ the\ vaccum\ (or\ a\ constant\ medium)}.\eqno (13)$$
In a varying medium where $V_\alpha$ and $V_\beta$ are also varying,
the neutrino encounters a resonance when
$$\delta m_M^2=0\quad {\rm or }\quad \delta m^2=2p(V_\alpha-V_\beta).\eqno
(14)$$
Then [19]
$$P_{\alpha\alpha}={1\over 2}+({1\over 2}-P_{jump})cos2\theta_{Mi}
cos2\theta_{Mf}\eqno (15)$$
where to linear order
$$P_{jump}={\rm exp}\Bigl[{-\pi\delta m^2sin^22\theta(V_\alpha-V_\beta)
\over 4cos2\theta}\Bigl({d\vert p(V_\alpha-V_\beta)\vert\over
dr}\Bigr)^{-1}\Bigl\vert_{res}\Bigr],\eqno (16)$$
where $r$ is the position of the neutrino.
We put the neutrino momentum $p$ inside the derivative because in cases
such as the expanding universe, $p$ changes with time.
When $P_{jump}\ll 1$ (``adiabtic''),
because $cos2\theta_{Mi}$ and $cos2\theta_{Mf}$ are of opposite sign on
each side of the resonance (at resonance, $\theta_M=45^\circ$),
the neutrinos mostly change their flavor;
while if $P_{jump}\sim 1$ (``nonadiabtic''),
the neutrinos mostly retain their initial flavor.
Therefore $P_{jump}$ serves as a indicator of the significance
of the resonance.

\centerline{\bf III.
Effects of Neutrino Oscillation on Primordial Nucleosynthesis}

The agreement of the predictions of Big Bang primordial nucleosynthesis
with the observed light element abundances has been one of the great
successes in Big Bang cosmology [21].
Therefore primordial nucleosynthesis has become an indispensable
probe of the early universe, and thereby tests theories of cosmology
and particle physics.
For example, it sets the best-known limit on the baryon to photon
ratio $\eta$ (=$N_B/N_\gamma$ as defined previously)
or equivalently the baryon density, $\rho_B$ [21]:
$$2.8\times 10^{-10}<\eta <4\times 10^{-10}\eqno (17)$$
It also rigorously constrains the total
number of light neutrino species ($m\la 1$ MeV), $N_\nu$,
that is in equilibrium at $T\sim$ MeV [21,22]:
$$N_\nu<3.6. \eqno (18)$$
Eq. (18) has been essentially confirmed
by the result of the $Z^0$ decay experiments at LEP, $N_\nu=2.99\pm 0.05$ [23].
The constraints from BBN
and that from LEP are complimentary because the latter only limits the
active neutrino species and the former limits both active and
sterile neutrino species [24].

What constrains the number of light neutrino species is the observed
$^4$He abundance $Y=0.235\pm 0.01$ [25].
Assuming only three known neutrino species,
standard Big Bang nucleosynthesis yields $0.236<Y<0.243$ [21].
When extra neutrino
species (therefore extra degrees of freedom) are introduced at the epoch of
$T\sim $ MeV in the early universe, the expansion of the universe would
be faster due to a higher energy density, which in turn leads to a higher
neutron to proton ratio when the ratio freezes out ($T\approx 0.7$ MeV),
and thus a higher helium yield [24]. Therefore the constraint of eq. (18)
can be obtained by requiring the calculated $Y$ assuming extra species
not to exceed the observations. An extra neutrino species, if it exists,
must interact weakly enough with the $Z^0$ to accomodate the LEP result.
Furthermore, it has to interact weakly enough so that it will not
be counted as one full species at the primordial nucleosynthesis epoch [24].
Therefore, a sterile neutrino might be allowed.

A sterile neutrino, i.e., a gauge group singlet, is
the simplest extension of the standard model [1].
It cannot be generated thermally in the epoch of nucleosynthesis
due to its sterility.
However, if it mixes with active neutrinos, the
active-sterile neutrino oscillation would produce sterile neutrinos
in the early universe, especially when MSW resonances were present.
The intensive production of sterile neutrinos through oscillation
or resonance before electron neutrinos decouple
may bring one more light neutrino species into chemical equilibrium,
and thus exceed the bound from the nucleosynthesis.
Furthermore, in the case of $\nu_e$-$\nu_s$ oscillation,
if the mass production of $\nu_s$ occurs after $\nu_e$ freezes out
but before the freeze-out of the neutron to proton ratio, the deficit
in $\nu_e$ relative to that counted as a full neutrino species
will cause the ratio to freeze out earlier and increase the $^4$He yield.
Therefore the mixing parameters between the sterile and active neutrinos
have to be constrained so that such
an intensive production of sterile neutrinos would not occur.

In the case of $\nu_e$ and $\nu_\mu$ (or $\nu_\tau$) mixing,
the neutrino oscillation can only show its signature on primordial
nucleosynthesis if there is number density asymmetry around $T\sim$ MeV
between $\nu_e$ and the species it oscillates into.
Within the framework of the standard model, a number density asymmetry
between $\nu_e$ and other active neutrino species can indeed be generated
at the 10$^{-2}$ level
by the larger branching ratio of $e^\pm$ annihilating into $\nu_e\bar\nu_e$
relative to other neutrinos [26]. Then one expects that a resonant
oscillation between $\nu_e$ and another active species
but not between their antiparticles, or vice versa,
will interchange their number densities
and lead to an asymmetry between $\nu_e$ and $\bar\nu_e$ [18].
This process doesn't change the total energy density of the universe,
so the expansion rate is unchanged.
But it changes the rates of weak interactions
that interchanges neutrons and protons. More $\nu_e$ than $\bar\nu_e$
leads to a higher rate for neutrons to be converted into protons, and
vice versa. However, as we will show later,
in the case of an oscillation between $\nu_e$ and another
active species, if the lepton number asymmetry is negligible,
$\nu_e$ and $\bar\nu_e$ will encounter resonances simultaneously
and the asymmetry between them will not be produced.
Furthermore, the asymmetry between the number densities of
$\nu_e$ and the other two active species may be washed out
in the case of large $\nu_e$-active neutrino oscillation.

The formalism of evolving the light
neutrino ensemble with neutrino oscillations
in the early universe has been carried out by various groups [15-17,27].
We will adopt the formalism of Enqvist {\sl et al.} [15].
At temperatures below 100 MeV and before the primordial
nucleosynthesis epoch, the universe was composed of
photons, electron-positrons, and neutrinos. (Nucleons are
present only at the 10$^{-10}$ level relative to other particle species.
Their contribution to the evolution of the neutrino ensemble is negligible.)
A natural hierachy in interation rates exist among these particle species.
Photons and electron-positrons interact via the electromagnetic interaction,
they reach equilibrium much faster than the neutrinos which
only interact weakly. Therefore in the temperature
region we are concerned with,
we can always assume photons and electron-positrons are in chemical
equilibrium. The neutrinos, however, get out of chemical equilibrium
at temperatures of several MeV (about 3 MeV for $\nu_e$ and $\bar\nu_e$,
5 MeV for $\nu_\mu$ and $\nu_\tau$, and their antiparticles) [21].
If assuming a two-flavor oscillation between $\nu_\alpha$ and $\nu_\beta$,
the density matrix of weak-interacting species can be expressed in the block
diagonal form:
$$\rho_W=\rho_\nu\oplus\rho_{\bar\nu}\oplus\sum_{i=e,\nu
 's\atop i\not=\nu_\alpha,\nu_\beta}(\oplus n_i\oplus n_{\bar i})\eqno (19)$$
where
$$\rho_\nu=\left(\matrix{\rho_{\alpha\alpha}&\rho_{\alpha\beta}\cr
\rho_{\beta\alpha}&\rho_{\beta\beta}\cr}\right)\eqno(20)$$

In cases of active-sterile neutrino mixing,
evolution of the neutrino ensemble becomes very
simple due to the sterility of $\nu_s$.
By rewriting
$$\rho_\nu={1\over 2}P_0(I+{\bf P}\cdot{\bf \sigma}),\quad
\rho_{\bar\nu}={1\over 2}\bar P_0(I+{\bf \bar P}\cdot{\bf \sigma}),\eqno (21)$$
and taking a thermal average by assuming all species are in
thermal equilibrium, the evolution equation for $\nu_e$-$\nu_s$
oscillation is (analogous expressions arise for the antineutrinos) [15]:
$$\eqalignno{{d\over dt}{\bf P}=&\bigl\langle{\bf V}\bigr\rangle
\times {\bf P}+(1-P_z)({d\over dt}{\rm ln}P_0)\hat{\bf z}\cr
&-(D^E+D^I+{d\over dt}{\rm ln}P_0)(P_x\hat{\bf x}+P_y\hat{\bf y}),&(22)\cr}$$
$${d\over dt}P_0=\sum_{\alpha=e,\nu_\mu,\nu_\tau}
\bigl\langle\Gamma(\nu_e\bar\nu_e\rightarrow\alpha\bar\alpha)\bigr\rangle
(\lambda_\alpha n_\alpha n_{\bar\alpha}-n_\nu_e n_{\bar\nu_e}),\eqno (23)$$

$${d\over dt}n_{\nu_\mu}=\sum_{\alpha=e,\nu_e,\nu_\tau}
\bigl\langle\Gamma(\nu_\mu\bar\nu_\mu\rightarrow\alpha\bar\alpha)\bigr\rangle
(\lambda_\alpha n_\alpha n_{\bar\alpha}-n_\nu_\mu n_{\bar\nu_\mu}),\eqno (24)$$
and similarly for $n_{\nu_\tau}.$
Here $\lambda_\nu=1$, $\lambda_e=1/4$ and the brackets $\langle\ldots\rangle$
indicate an average over the momentum distribution of the ensemble.
$n_\alpha$ is the number density relative to that of a fully relativistic
fermion with single helicity. Therefore $n_e=2$ when $T\ga m_e$, and
$n_\nu$=1 if they are in chemical equilibrium. The contributions of
right-handed neutrinos are neglected since we assume the neutrinos are
light and they have only the weak interaction.

{\bf V} (and analogously for $\bar{\bf V}$) of $\nu_e$-$\nu_s$ mixing is
$${\bf V}=2\omega_{es}\hat{\bf x}+(V_e+\omega_{ee}-\omega_{ss})\hat{\bf y}
\eqno (25)$$
in which
$$2\omega_{es}={\delta m^2\over 2p}sin2\theta, \quad
\omega_{ee}-\omega_{ss}=-{\delta m^2\over 2p}cos2\theta.\eqno (26)$$
We assume the sterile neutrino is heavier than the active neutrino
if $\delta m^2>0$, and vice versa.

$D^I$ and $D^E$ are the damping coefficients due to inelastic and elastic
collisions of the active neutrinos
$$\eqalignno{D^E&={1\over 2}
\biggl[\bigl\langle\Gamma(\nu_e e^-\rightarrow\nu_e e^-)\bigr\rangle n_{e^-}
+\bigl\langle\Gamma(\nu_e e^+\rightarrow\nu_e e^+)\bigr\rangle n_{e^+}\biggr]
\cr &+{1\over 2}\sum_{\alpha=e,\mu,\tau}
\biggl[\bigl\langle\Gamma(\nu_e\nu_\alpha\rightarrow\nu_e\nu_\alpha)\bigr
\rangle n_{\nu_\alpha}
+\bigl\langle\Gamma(\nu_e\bar\nu_\alpha\rightarrow\nu_e\bar\nu_\alpha)\bigr
\rangle n_{\bar\nu_\alpha}\biggr]&(27)\cr}$$

$$D^I={1\over 2}
\biggl[\bigl\langle\Gamma(\nu_e\bar\nu_e\rightarrow e^+ e^-)\bigr\rangle
+\sum_{\alpha=\mu,\tau}
\bigl\langle\Gamma(\nu_e\bar\nu_e\rightarrow\nu_\alpha\bar\nu_\alpha)\bigr
\rangle \biggr]n_{\bar\nu_e}. \eqno (28)$$
$\bar D^I$ and $\bar D^E$ for antineutrinos can be deduced by interchanging
$\nu_e$ and $\bar\nu_e$.

The above formalism can be anologously inferred
for $\nu_\mu$-$\nu_s$ or $\nu_\tau$-$\nu_s$
oscillation by interchanging $\nu_\mu$ or $\nu_\tau$ with $\nu_e$.

The thermal averaged reaction rate $\Gamma$'s
has been calculated previously, e.g. in ref. 15 and ref. 26.
Table 1 lists their values in units of $F_0$, where
$$F_0={G_F^2\langle p\rangle^2\over 6\pi}N_\gamma(T)\eqno (29)$$
Since $T_\gamma$ and $T_\nu$ differ by at most 1$\%$ in the epoch
in which we are interested [26], from now
on we shall ignore their difference and denote them by $T$.

Before going ahead and solving eqs. (22)--(24) numerically, it is
helpful to do an analytical estimate of the problem.
Because $\nu_s$ is only produced through oscillation,
its production rate is approximated by
$$\eqalignno{\Gamma_{\nu_s}^{prod}&
\simeq{1\over 2}sin^22\theta_M\Gamma_{\nu}
\approx 2 G_F^2 T^5 sin^22\theta_M.&(30)\cr}$$
At high temperature the oscillation is suppressed
by the matter effect which is proportional to $T^6$.
The mixing angle at high temperature is,
$$\eqalignno{tan2\theta_M\approx -{sin2\theta
\over 2p(V_\alpha-V_\beta)/\delta m^2}&\sim sin2\theta\Bigl(
{5\times 10^{20}{\rm eV}^2\delta m^2\over G_F p^2T^4}\Bigr)
\quad{\rm for\ }\nu_e{\rm -}\nu_s {\rm\ mixing,}\cr
&\sim sin2\theta\Bigl(
{2\times 10^{21}{\rm eV}^2\delta m^2\over G_F p^2T^4}\Bigr)
\quad{\rm for\ }\nu_{\mu,\tau}{\rm -}\nu_s {\rm\ mixing.}&(31)\cr}$$
If one takes $p\sim 3T$, when
$$\eqalignno{T({\rm MeV})\ga &\bigl({\delta m^2\over 10^{-8}{\rm eV}^2}\bigr)
^{1/6},\quad{\rm for\ }\nu_e{\rm -}\nu_s {\rm\ mixing,}\cr
&\bigl({\delta m^2\over 10^{-9}{\rm eV}^2}\bigr)
^{1/6},\quad{\rm for\ }\nu_{\mu,\tau}{\rm -}\nu_s {\rm\ mixing,}&(32)\cr}$$
the mixing is suppressed at the $sin^22\theta_M\sim
10^{-4}sin^22\theta$ level so that no significant amounts of
$\nu_s$ can be produced.
Only when the temperature of the universe falls significantly
below this temperature, does the $\nu_s$ production occur noticably.
Therefore mixings with
$\delta m^2<10^{-8}$ to $10^{-9}$ eV$^2$ will never produce
appreciable amount of $\nu_s$'s before the weak freeze
out of baryons at $\sim 0.7$ MeV which affect primordial nucleosynthesis.

The $\nu_s$ will be brought into chemical equilibrium if
$\Gamma_{\nu_s}^{prod}$ is larger than the Hubble constant
before active neutrinos lose thermal contact with the ambient matter,
which occurs at $T\sim 1$ MeV.
For $\delta m^2>0$, $\Gamma_{\nu_s}^{prod}/H$ has a maximum at
$$T_{max}({\rm MeV})= B_{\nu_\alpha}
\bigl({\delta m^2\over {\rm 1 eV}^2}\bigr)^{1/6}\bigl({p\over T}\bigr)^{-1/3},
\eqno (33)$$
where $B_{\nu_e}\approx 7$ and $B_{\nu_{\mu,\tau}}\approx 9$.
The requirement that
$${\Gamma_{\nu_s}^{prod}\over H}\biggr\vert_{T_{max}}<1\quad
{\rm and}\quad T_{max}\ga 1{\rm MeV}\eqno (34)$$
gives
$$\delta m^2 sin^42\theta\ga 10^{-4}{\rm eV}^2\quad
{\rm for\ }\nu_e (\nu_\mu\ ,or\ \nu_\tau}){\rm -}\nu_s\
{\rm mixing}.\eqno (35)$$

For $\delta m^2<0$, a resonance occurs at
$$T_{res}({\rm MeV})=C_{\nu_\alpha}\biggl
({\vert\delta m^2cos2\theta\vert\over p^2/T^2}\biggr)^{1/6},\eqno (36)$$
where $C_{\nu_e}\approx 19$ and $C_{\nu_{\mu,\tau}}\approx 23$.
The transition probability $P_{jump}$ has to be sufficiently small so that
sizable numbers of $\nu_s$ can be converted from active neutrinos.
A rough requirement is that $P_{jump}< e^{-1}$. Therefore
$$\Bigl({\pi\delta m^2sin^22\theta\over 4cos2\theta}\Bigr)
\biggl\vert {6p\over T}{dT\over dt}\biggr\vert ^{-1}_{T_{res}}>1\quad
{\rm and}\quad T_{max}\ga 1 {\rm MeV},\eqno (37)$$
from which we get
$$\delta m^2 sin^42\theta\ga 10^{-9}{\rm eV}^2\quad
{\rm and}\quad \delta m^2\ga 10^{-7}{\rm eV}^2\eqno (38)$$
for all three oscillation cases.

When the major production of $\nu_s$ through $\nu_e$-$\nu_s$ oscillation
occurs after the thermal
freeze out of $\nu_e$ and before the weak freeze out of baryons,
additional limits can be obtained.
The resultant $\nu_e$ deficit relative to a full relativistic species,
 $\delta n_{\nu_e}$ (defined to be positive if it is a deficit),
will increase the neutron to proton ratio, and hence the $^4$He
yield. The constraint that $N_\nu<3.6$ is replaced approximately by [15]
$$N_\nu+5\delta n_{\nu_e}<3.6.\eqno (39)$$
In this case, $\nu_\mu$ and $\nu_\tau$ are still approximately
two full relativistic species. $\nu_e$ plus $\nu_s$
are roughly counted as one. Therefore $N_\nu\approx 3$.
In the case of $\delta m^2>0$,
The deficit $\delta n_{\nu_e}$ is roughly $0.5sin^22\theta_M$.
Hence mixings with
$$\eqalignno{&sin^22\theta \ga 0.25\quad{\rm for}\quad
\delta m^2>10^{-7}{\rm eV^2}\cr
&\delta m^2sin^22\theta \ga 10^{-8}\quad{\rm for}\quad
\delta m^2<10^{-7}{\rm eV^2}&(40)\cr} $$
are then excluded by eq. (39).
For $\delta m^2<0$, the requirement that the resonance occuring
between 1 MeV and 0.7 MeV has to be sufficiently nonadiabatic to satisfy
eq. (39), sets a stronger limit; $\delta n_{\nu_e}$ is roughly $1-P_{jump}$.
Therefore $\delta n_{\nu_e}\la 0.12$, which translates into
$$\delta m^2 sin^42\theta\ga 10^{-10}{\rm eV}^2\quad
{\rm for}\quad &10^{-8}{\rm eV}^2\la\delta m^2\la 10^{-7}{\rm eV}^2.\eqno
(41)$$

Our results in eqs (35), (38), (40) and (41) are similar to the result of
Enqvist {\sl et al.} [15] and Cline [17]. It should be
noted that the $\Gamma_\nu$ in eq. (30) is
the total scattering rate of active neutrinos, instead of only
the inelastic scattering rate as adopted by  Barbieri {\sl et al.} [16].
Elastic scatterings reduce the mixed neutrino states
into either pure active neutrino states or pure sterile neutrino states
(just as a polarizer reduces a beam of light into either parallel polarized
or perpendicularly polarized).
The pure active neutrinos will develope sterile components via oscillations.
(It is also true that the pure sterile neutrinos
will oscillate into mixed states.
But if $sin^22\theta_M$ is small, the active components
in the resultant mixed states are negligible.)
The net effect is the production of sterile neutrinos through
elastic scatterings of neutrinos.

Figures 1 to 4 show the constant
helium yield, $Y$, contours from numerical integrations of
eq. (22) to (24) in different cases (with $\eta=2.8\times 10^{-10}$
which gives the most conservative $Y$ estimates [22]). Fig. 1 also shows
the $\nu_e$-$\nu_s$ MSW mixing solution
to the solar neutrino problem at $95\%$ C. L. (using a Monte Carlo
calculation of 1000 solar models similar to that in ref. 12, which
represents the astrophysical uncertainties in the solar models).
The $\nu_\mu$-$\nu_s$ mixing solution
to the atmospheric neutrino problem is shown in Fig. 3 and 4.
Parameter spaces on the upper right
of the $Y=0.245$ contour yield a higher $^4$He abundance
than found by observation, and thus
are excluded. {\sl
The excluded regions include the large angle $\nu_e$-$\nu_s$ MSW
solution to the solar neutrino problem and the $\nu_\mu$-$\nu_s$ mixing
solution to the atmospheric neutrino solution}.
We express the constraints in terms of the primordial helium yield $Y$ instead
of $N_\nu$ as in ref. 15 because $N_\nu$
gradually changes in the course of the early universe. It is the final
helium yield that is to be directly compared with observation.
In the integration of eqs. (22)--(24), we assumed particle-antiparticle
symmetry, i.e., $L_\alpha\equiv 0$. The justification of such an
assumption has been investigated by Enqvist {\sl et al.} [28].
They concluded that as long as
there is no resonance, an initial $CP$ asymmetry will be driven toward
$L_\alpha=0$, which is then a fixed point of eqs. (22) to (24).
With a large lepton number asymmetry (e.g., larger than 10$^{-7}$),
the first terms in the effective potentials eq. (7) and (8) would dominate.
Constraints would then be modified accordingly. Nonetheless,
The large angle mixing ($sin^22\theta\sim 1$) between sterile neutrinos
and active neutrinos could still be confidently ruled out
for $\delta m^2\ga \bigl({L\over 10^{-7}}\bigr) 10^{-6}{\rm eV}^2$, where
$L$ is the lepton asymmetry.

Another concern is that eqs. (22)--(24) are only accurate as long as
the neutrinos maintain a thermal spectrum.
Without the oscillations, the neutrino ensemble will have a
thermal spectrum through out the history of primordial nucleosynthesis
if one neglects the $e^\pm$ heating at the $10^{-2}$ level.
Possible thermal distortion may be generated from those neutrinos
with different momentum that begin to
oscillate into $\nu_s$ at different temperatures with
different amplitudes, and the weak interaction cannot fully compensate
for the difference. However, we will see that such oscillations
are not very sensitive to the neutrino momentum, and as a result, they don't
generate significant spectral distortions.

When the oscillation occurs before the freeze out of active neutrinos
and $\delta m^2>0$, we can
see from eq. (33) that the maximal oscillation temperature depends weakly
on momentum, $T_{max}\propto p^{-1/3}$. Moreover, $\Gamma_s^{prod}$
is always less than the weak interaction rates of active neutrinos.
Therefore the weak interactions can always keep the active neutrinos
in thermal contact.
When $\delta m^2<0$, It is also shown that $P_{jump}$
is independent of $p$, while $T_{res}\propto p^{-1/3}$ [15].
Therefore, the possible spectral distortion can only be a higher order effect.

When the large numbers of oscillations between $\nu_e$ and $\nu_s$
occur after the chemical freeze out
of $\nu_e$, a possible distortion occurs due to the dependence of the magnitude
of the oscillation $0.5sin^22\theta_M$ on momentum.
When $\delta m^2\ga 10^{-7}$eV$^2$, the matter effect is small
compared to the vaccum mixing, $sin^22\theta_M\approx sin^22\theta$,
and the thermal distribution assumption is still valid.
Below $10^{-7}$ eV$^2$, we are only excluding very large angle mixings and
the $p$ dependent matter effect is not of major concern.

Overall, for the mixing parameters we probe, a thermal
distribution is a reasonable approximation with which
to evolve the neutrino ensemble.
The effect of a possible nonthermal distribution may be
estimated by inserting different $p$'s into {\bf V} instead of
a thermal average. We found that the constraint is not sensitive to
the $p$ value we put in, which confirms the validity of a thermal
distribution.

In cases of active-active neutrino oscillation,
the evolution of the neutrino ensemble is much more
complicated to calculate. In the standard Big Bang cosmology,
the number densities of $\nu_\mu$ and $\nu_\tau$ are identical,
and the number density of $\nu_e$ is slightly higher (at the 10$^{-2}$ level
at MeV temperatures) than the other two species due to the charged current
heating of the electron positron annihilations [26]. Therefore the introduction
of a $\nu_\mu$-$\nu_\tau$ mixing will have no effect on
Big Bang nucleosynthesis. The effect of a $\nu_e$-$\nu_\mu$
mixing (the same with $\nu_e$-$\nu_\tau$ oscillation)
has been estimated previously [18]. But a careful analysis and calculation
has yet to be done. It was argued that in the case of a large
lepton asymmetry in the universe, a resonance between $\nu_e$
and $\nu_\mu$, if sufficiently adiabatic and occuring between
$\sim 0.7$ MeV and $\sim 3$ MeV, will swap the number densities
between the $\nu_e$ and $\nu_\mu$ but not
$\bar\nu_e$ and $\bar\nu_\mu$ (or vice versa),
and lead to a small deficit in $\nu_e$ with respect to $\bar\nu_e$,
and therefore change the neutron to proton ratio [18].
However, for a small lepton asymmetry at the $10^{-9}$ level,
the finite temperature terms dominate the effective potentials in
eq. (7) and (8) at $T\ga 1$ MeV. Therefore the resonance between
$\nu_e$ and $\nu_\mu$ and the resonance between $\bar\nu_e$ and $\bar\nu_\mu$
occur almost simultaneously with similar strength and no
sizable asymmetry between $\nu_e$ and $\bar\nu_e$ can be produced,
even at the 10$^{-2}$ level.

We further found that in the presence of a mixing between $\nu_e$
and $\nu_\mu$, the existance of such an
asymmetry between the number densities of $\nu_e$ and $\nu_\mu$,
is questionable. Using the same formalism as in the eqs. (21),
this aymmetry corresponds to a non-zero positive $P_z$.
The rate of increasing $P_z$, i.e., the rate of generating the
asymmetry is roughly $(\Gamma_{e^-e^+\rightarrow\nu_e\bar\nu_e}-\Gamma_{e^-e^+
\rightarrow\nu_\mu\bar\nu_\mu})\Delta n_e$, where $\Delta n_e$
is the number density
of electrons annihilated with respect to a fully relativistic
species; At $T\sim$ MeV, $\Delta n_e\sim 0.1$. The rate of damping $P_z$
by the weak interactions between the flavor eigenstates
is roughly 0.5$(\Gamma_{\nu_e}+\Gamma_{\nu_\mu})sin^22\theta_M$,
where $\Gamma_{\nu_e}$ and $\Gamma_{\nu_\mu}$ are the rates of the
$\nu_e$ and $\nu_\mu$'s weak interaction.
As long as the damping rate is larger than the production rate,
$P_z$ will remain 0. Since $(\Gamma_{\nu_e}+\Gamma_{\nu_\mu})> 10^2
(\Gamma_{e^-e^+\rightarrow\nu_e\bar\nu_e}-\Gamma_{e^-e^+
\rightarrow\nu_\mu\bar\nu_\mu})\Delta n_e$, to keep $P_z$ at zero
requires
$$sin2\theta_M\ga 10^{-1}\quad {\rm at}\ T\sim{ MeV}.\eqno (42)$$
Therefore for mixings with
$$\delta m^2\ga 10^{-7}{\rm eV}^2\quad {\rm and}\quad sin2\theta\ga 10^{-1},
\eqno (43)$$
the number density asymmetry between
$\nu_e$ and $\nu_\mu$ will be damped to a much lower level than 10$^{-2}$.
The ``miraculous'' cancellation mentioned in ref. 26 would then
no longer exist. The net change in $^4$He yield with respect to
standard Big Bang nucleosynthesis considered in ref. 26 would then
be $\sim -10^{-4}$ to $-10^{-3}$ instead of $\sim 10^{-5}$.
The large angle $\nu_e$-$\nu_\mu$ MSW mixing solution to the
solar neutrino problem clearly belongs to this case,
but a change in the helium abudance at the $10^{-3}$ level can hardly
put any constraints on the mixing parameter space of
$\nu_e$-$\nu_\mu$ oscillations.
\centerline{\bf IV. The Solar Neutrino Implications}
{}From Fig. 1 we can see that the large angle $\nu_e$-$\nu_s$ MSW mixing
solution to the solar neutrino problem  is clearly ruled out.
It has been a concern before that the next generation solar neutrino
experiments may not unambigously resolve various
new neutrino physics solutions from a possible solar model solution to
the solar neutrino problems. In particular, the large angle $\nu_e$-$\nu_s$
MSW mixing and modified solar model solutions all fail to exhibit
spectral distortions with respect to the $\beta$-decay spectra
nor do they show enhanced neutral current events.
Now that the large angle $\nu_e$-$\nu_s$ MSW mixing solution has been excluded
by primordial nucleosynthesis, this ambiguity is removed.
Two major next generation solar neutrino experiments, the Sudbury Neutrino
Observatory (SNO) and Super-Kamiokande (Super-K) [2], both planning to start
operation in 1996, will be sensitive to spectral distortions with
respect to the $\beta$-decay spectra. Furthurmore, SNO will be capable of
measuring neutral current events and hence identify
$\nu_e$-$\nu_{\mu}$ (or $\nu_\tau$) oscillations.
With the operation of these two experiments, as well as other next generation
solar neutrino experiments (Borexino, ICARUS, etc., [2]),
one may in principle resolve the different solutions to the solar neutrino
problem, in particular, the conflict between new neutrino physics solution
and the modified solar model solution (which is currently still allowed
if there is a problem with the chlorine experiment) [6].
In table 2, we list the signatures of six different solutions in
next generation experiments. All $\nu_e$-$\nu_\mu$ (or similarly $\nu_\tau$)
mixing solutions will
exhibit extra neutral current events in SNO, due to the neutral current
scattering of $\nu_\mu +{\rm d}\rightarrow\nu +{\rm p}+{\rm n}$.
The small angle MSW mixing between $\nu_e$-$\nu_\mu$
or $\nu_e$-$\nu_s$, will exhibit significant spectral distortion with
respect to a $\beta$-decay spectrum. Part of the large angle
$\nu_e$-$\nu_\mu$ MSW mixing parameter space
also exhibits spectral distortion to some
degree, when $sin^22\theta\sim 0.5$. But when $sin^22\theta$ is close to 1,
where most of the large angle mixing solution lies,
the spectrum is more or less uniformly suppressed with respect to the
normal $\beta$ spectrum [10].
The vacuum mixing will also distort the neutrino spectrum,
due to its strong energy dependence. Furthermore, its
strong distance dependence leads to uniquely
large seasonal variation in the observed $^7$Be neutrino flux
(besides the usual $1/R^2$ dependences of neutrino flux),
which can be detected in the Borexino experiment [10].
The modified solar model solutions, on the other hand, are immune to
all the above anomolies.
The resonant spin precession solution (based on neutrino magnetic moments)
is not very well understood since
the magnetic field in the sun is largely unknown. But,
it is not strongly viable due to the bounds on the solar magnetic field
and the need for a large
neutrino magnetic moment. However, one also expects spectral distortions
and 11-year variation
since the adiabaticy of the resonance is sensitive to the neutrino
energy, and the field configuration [29].

\centerline{\bf V. Dark Matter Implications}
It is interesting to attempt to
relate three astrophysical and cosmological problems
that involve neutrinos: the solar neutrino problem,
the atmospheric neutrino problem, which were discussed above,
and the cosmological dark matter problem.
Big Bang nucleosynthesis constrains the baryon density to be less
than 10$\%$ of the critical density. The remaining 90$\%$ of the mass
must be non-baryonic dark matter [30], assuming a critical density
universe. One possible candidate for dark matter is
a 30 eV neutrino species (e.g., $\nu_\tau$), Hot Dark Matter (HDM) model [31].
However, HDM with adiabatic density fluctuations
fails to explain structures on small scales of the universe
($\sim $Mpc). Its alternative, the Cold Dark Matter (CDM) with
adiabatic density fluctuations lacks power
on large scales to simultaneously fit the data from COBE,
APM and IRAS well when normalized to the galaxy distributions at 10 Mpc [32].
It was suggested recently that a reasonable fit to the data
is a cocktail of a 7 eV neutrino and some cold dark matter [33].
Therefore, to consider a $\sim 10$ eV neutrino
as the dark matter of the universe is still well motivated.
Also, pure HDM with topological defects as seeds still
seems to work reasonably well [34].

It has been known [35] that with only 3 generation of neutrinos and
with the most natural particle models that generate
neutrino masses, namely the see-saw mechanism,
it is difficult to simultaneously reconcile the three
astrophysical and cosmological problems.
The solar neutrino deficit observed by
the current solar neutrino experiments suggests a mixing between
$\nu_e$ and another species $\nu_x$ with $m_{\nu_x}^2-m_{\nu_e}^2$
of order 10$^{-6}$ to 10$^{-4}$ eV$^2$ (or $10^{-10}$ eV$^2$ in the
case of vaccum mixing), While the $\nu_\mu$ deficit, if real, in the
atmospheric neutrino experiments, suggests a large mixing between
$\nu_\mu$ and $\nu_y$ with $\vert m_{\nu_\mu}^2-m_{\nu_y}^2\vert$
of the order 10$^{-3}$ to 10$^{-1}$ eV$^2$.
If $\nu_x$ and $\nu_y$ were active neutrinos, i.e.,
$\nu_\mu$ or $\nu_\tau$, one immediately sees that, unless the three masses
are almost degenerate, (which may not be natural theoretically,)
no neutrino species can provide a $\sim$10 eV mass to serve as dark matter.
The see-saw mechanism
predicts a mass hierachy $m_{\nu_e}\ll m_{\nu_\mu}\ll m_{\nu_\tau}$ [36].
Therefore it seems that a sterile neutrino $\nu_s$
must be introduced to reconcile the three astrophysical/cosmological
problems without conflicting with the constraint from LEP that only three
active neutrino species exist [23]. There are three sterile neutrino options
(assuming all three problems are real and require neutrino solutions):

1. $\nu_x$ is the sterile neutrino $\nu_s$, and $\nu_y$ is $\nu_\tau$.
The matter mixing or the vaccum mixing
of $\nu_e$ and $\nu_s$ explains the solar neutrino deficit.
Our calculation above has excluded the large mixing
angle MSW solution. The small angle solution or the vaccum solution
will then be the solution to
the solar neutrino problem, and will show very distinct features in
future solar neutrino experiments, as seen from Table 2.
The $\nu_\mu$-$\nu_\tau$ mixing with $\delta m^2\sim
10^{-3}$ to $10^{-1}$ eV$^2$ implies $m_\mu\approx m_\tau\sim 10$ eV.
Models with this prediction require
$\nu_\mu$ and $\nu_\tau$ to be Zeldovich-Konopinski-Mahmoud (ZKM) type
neutrinos [1].

2. $\nu_x$ is $\nu_\mu$ and $\nu_y$ is $\nu_s$. Then in order to solve the
atmospheric neutrino problem, a large angle mixing between $\nu_\mu$ and
$\nu_s$ is needed, as in Fig. 3 and 4. This clearly conflicts with
the nucleosynthesis bound we obtained and is no longer viable.

3. The final option is that $\nu_x$ and $\nu_y$ are $\nu_\mu$ and $\nu_\tau$
respectively, and $\nu_s$ provides the dark matter.
As seen above, $\nu_s$ can be produced by mixing
with active neutrinos in the early universe. With a mass of $\sim 10$ eV or
more, the $\nu_s$ can serve as the dark matter without conflicting with the
nucleosynethesis bound, given appropriate mixing angles.
The energy density $\Omega_{\nu_s}$ of $\nu_s$ relative to
the critical density, is roughly [37]
$$\Omega_{\nu_s}h^2\sim{\Gamma_{\nu_s}^{prod}\over H}\biggr\vert_{T_{max}}
\bigl({m_{\nu_s}\over 30{\rm eV}}\bigr),\eqno (44)$$
where $h=H_0/$(100km/sec/Mpc) and $H_0$ is Hubble constant today.
To have $\Omega_{\nu_s}\sim 1$ requires
$$m_{\nu_s}sin2\theta\sim 0.3h\ {\rm eV}\quad {\rm and}\quad
10{\rm eV}\la m_{\nu_s}\la 1 {\rm keV} \eqno (45)$$

The cap on $m_{\nu_s}$ comes from requiring
$T_{max}$ in eq. (33) to be less than the mass of muons or
the temperature of the quark-hadron
phase transition, which is $\sim$100 MeV, so that the total number
of degrees of freedom (excluding the produced sterile neutrinos)
is roughly a constant during the oscillation, and the previous formalism
applies [24].
For a sterile neutrino with a mass larger than 1 keV, large amount of
$\nu_s$ production through the oscillation occurs before the quark-hadron
phase transition.
Since the degrees of freedom before the quark-hadron phase transition
are $\sim 10^{2}$ vs. $\sim 10$ after, the sterile neutrino population
would be significantly diluted. Therefore the presence of
such a neutrino as a full species before the quark-hadron phase
transition will not exceed the nucleosynthesis bound.
The mixing parameters that
provide $\Omega_{\nu_s}\sim 1$ are not simple to calculate due to
the complication of the quark-hadron transition and the presence of muons,
pions and quarks.

As we can see, there exists a
range of sterile neutrino masses and mixing angles
that can provide a condidates for either hot dark matter models or
cocktail models (consisting of both hot dark matter matter and cold
dark matter). Also, compared with conventional $\sim 10$ eV
active neutrino dark matter, a heavier sterile neutrino is allowed
and has a smaller free streaming length $\lambda_{FS}$.
In other words it won't damp out fluctuations
on as small a scale as a 10 eV neutrino would do. Since [30]
$$\lambda_{FS}\sim 20{\rm Mpc}\bigl({m_\nu\over 30{\rm eV}}\bigr)^{-1},
\eqno(46)$$
several hundred eV sterile neutrino dark matter [37] will
then have a free streaming length of $\sim$
1 Mpc, which is the scale of a galaxy,
and might allow neutrinos to serve as dark matter even with adiabatic density
fluctuations.

\centerline{\bf VI. Summary}
In summary, we examined the effect of neutrino mixings in Big
Bang Nucleosynthesis, and calculated the constraints from the primodial
$^4$He abundance on the mixing parameters of sterile-active neutrino mixings.
We discussed the implications of these constraints on the solar neutrino
problem, the atmospheric neutrino problem, and the cosmological
dark matter problem. We conclude that future solar neutrino experiments may
unambiguously differentiate solutions to the solar neutrino problem,
in particular, the new neutrino physics solutions and solar model solutions.
We also conclude that a $\nu_\tau$ of $\sim 10$ eV or
a sterile neutrino of 10 eV to 1 keV
with proper mixing with active neutrinos may provide the cosmological
dark matter, depending on whether $\nu_e$-$\nu_s$ mixing or
$\nu_e$-$\nu_\mu$ mixing solves the solar neutrino problem.

\centerline{\bf Acknowledgement}
We thank S. Dodelson for intriguing discussions about the role of
sterile neutrinos as dark matter. We also thank J. N. Bahcall for providing
us with the 1000 Monte Carlo selected solar models.
This work was supported by DoE (Nuclear) and
NSF at the Unversity of Chicago, and by the DoE and by NASA through
grant NAGW 2381 at Fermilab.
\vfill\eject

\noindent {\bf Reference:}

\noindent [1] P. Langacker, UPR-0511T, presented at {\sl Neutrino Physics},
Venice, March 1992, and references therein.

\noindent [2] John N. Bahcall, {\sl Neutrino Astrophysics},
Cambridge University Press (1989).

\noindent [3] R. Davis, Jr., {\sl et al.}, in {\sl Proceedings of the 21th
International Cosmic Ray Conference}, 1990, V 12, edited by R. J. Protheroe
(University of Adelaide Press, Adelaide), p. 143.

\noindent [4] S. Turck-Chi\`eze, S. Cahn, M. Cass\'e, and C. Doom,
{\sl Ap. J.}, {\bf 335}, 415 (1988); S. Turck-Chi\`eze and I. Lopes,
{\sl Ap. J.}, 1992 (to be published).

\noindent [5] T. K. Kuo and James Panteleone, {\sl Rev. Mod. Phys.}
 {\bf 61}, 937 (1989).

\noindent [6] X. Shi and D. N. Schramm, {\sl Particle World},
1993, in press.

\noindent [7] K. S. Hirata {\sl et al.}, {\sl Phys. Lett.} {\bf B 250},
146 (1992).

\noindent [8] D. Casper, {\sl et al.}, {\sl Phys. Rev. Lett.}
{\bf 66}, 2561 (1991).

\noindent [9] P. Litchfield, in {\sl Proceedings of the International
Workshop on $\nu_\mu/\nu_e$ Problem in Atmospheric Neutrinos},
eds. V. Berezinsky and G. Fiorentini, March 1993, Italy.

\noindent [10] P. I. Krastev, S. T. Petcov, {\sl Phys. Lett.} {\bf 299B},
99 (1992); CERN-TH.6539/92.

\noindent [11] L. Wolfenstein, {\sl Phys. Rev.} D {\bf 17}, 2369 (1978);
S. P. Mikheyev and A. Yu. Smirnov, {\sl Sov. J. Nucl. Phys.} {\bf 42}, 913
(1985).

\noindent [12] X. Shi, D. N. Schramm and J. N. Bahcall,
{\sl Phys. Rev. Lett.}, {\bf 69}, 717 (1992).

\noindent [13] J. G. Learned, S. Pakvasa, and T. J. Weiler,
{\sl Phys. Lett.} B {\bf 207}, 79 (1988);
V. Barger and K. Whisnant, {\sl Phys. Lett.} B {\bf 209}, 365 (1988).

\noindent [14] A summary can be found in, G. Bernardi, {\sl XXIV Int.
Conf. on High Energy Physics,} ed. R. Kotthaus and J. H. K\"uhn,
p. 1076 (Springer, Berlin, 1989).

\noindent [15] Enqvist {\sl et al.}, {\sl Nucl. Phys.} {\bf B373}, 498 (1992).

\noindent [16] R. Barbieri and A. Dolgov, {\sl Nucl. Phys.}, {\bf B 349},
743 (1991).

\noindent [17] J. M. Cline, {\sl Phys. Rev. Lett.}, {\bf 68},
3137 (1992).

\noindent [18] P. G. Langacker {\sl et al.}, {\sl Nucl. Phys.},
 {\bf B 266}, 669 (1986); M. J. Savage {\sl et al.}, {\sl Ap. J.},
{\bf 368}, 1 (1991).

\noindent [19] T. K. Kuo and James Panteleone, {\sl
Rev. Mod. Phys.}, {\bf Vol. 61}, No. 4, 937 (1989).

\noindent [20] D. N\"otzold and G. Raffelt, {\sl Nuclear Physics} {\bf B 307},
924 (1988).

\noindent [21] T. P. Walker, G. Steigman, D. N. Schramm, K. Olive
and H.-S. Kang, {\sl Ap. J.} {\bf 376}, 51 (1991).

\noindent [22] T. P. Walker, {\sl Ap. J.}, 1993, in press; OSU-TA-1/93;
D. N. Schramm and L. Kawano,
{\sl Nuc. Instruments and Meth.} {\bf A284}, 84-88 (1989).

\noindent [23] For a review, see J. R. Carter, in  {\sl Proceedings of the
Lepton Photon and High Energy Physics Conference}, Geneva, July 1991.

\noindent [24] K.A. Olive, D.N. Schramm and G. Steigman.
{\it Nucl. Phys.} {\bf 180}, 497-515 (1981).

\noindent [25] E. Skillman {\sl et al.}, {\sl Ap. J.}, 1993, in press;
also, in {\sl Proc. 16th TEXAS/PASCOS Meeting}, Berkeley, 1992.

\noindent [26] S. Dodelson and M. S. Turner, {\sl Phys. Rev.} {\bf D 46},
3372 (1992).

\noindent [27] A. Dolgov, {\sl Sov. J. Nucl. Phys.}, {\bf 33}, 700 (1981);
L. Stodolsky, {\sl Phys. Rev.} {\bf D 36}, 2273 (1987).

\noindent [28] K. Enqvist {\sl et al.}, {\sl Nuclear Physics}, {\bf B 349},
754 (1991).

\noindent [29] C.-S. Lim and W. J. Marciano, {\sl Phys. Rev.}
D {\bf 37}, 1368 (1988); X. Shi {\sl et al.}, {\sl Comments on Nuclear
and Particle Physics}, 1993, in press.

\noindent [30] E. W. Kolb and M. S. Turner, {\sl The Early Universe},
(Addison-Wesley, 1990);
E. Kolb, D. Schramm and M. Turner, in {\sl Neutrino Physics},
ed. K. Winter, (Cambridge University Press, 1991).

\noindent [31] K. Freese and D.N. Schramm.
{\it Nuclear Physics} {\bf B233}, 167 (1984).

\noindent [32] D. N. Schramm, in {\sl Proc. Takayama Neutrino Phys.
Meeting}, 1993, and references therein.

\noindent [33] Marc Davis {\sl et al.}, {\sl Nature}, {\bf V 359}, 393 (1992).

\noindent [34] Andreas Albreicht and Albert Stebbins, {\sl Phys. Rev. Lett.}
{\bf 69}, 2615 (1992).

\noindent [35] S. A. Bludman {\sl et al.}, {\sl Nucl. Phys.} {\bf B, 374},
373 (1992).

\noindent [36] M. Gell-Mann, P. Ramond, R. Slansky, in {\sl Supergravity},
 eds. P. van Nieuwenhuizen and D. Freeman (North-Holland, 1979).

\noindent [37] A detailed calculation can be found in S. Dodelson and
L. M. Widrow, FERMILAB-Pub-93/057-A; hep-ph-9303287.

\vfill\eject
\noindent{\bf Figure captions:}

\noindent Fig. 1. A contour plot of constant primordial helium yield contours
for $\nu_e$-$\nu_s$ mixing with $\delta m^2>0$.
The parameter space to the right of the $Y=0.245$ contour is excluded
by the current observations. The area enclosed by dashed lines is the MSW
solution to the solar neutrino problem at 95$\%$ C. L.,
with astrophysical uncertainties taken into account.

\noindent Fig. 2. The same contour plot as in Fig. 1 for
$\nu_e$-$\nu_s$ mixing with $\delta m^2<0$.

\noindent Fig. 3. The same contour plot as Fig. 1 for
$\nu_\mu$-$\nu_s$ mixing with $\delta m^2<0$. The area enclosed by
the dashed line is the solution to the atmospheric neutrino problem.

\noindent Fig. 4. The same contour plot as Fig. 3 for
$\nu_\mu$-$\nu_s$ mixing with $\delta m^2>0$.

\vfill\eject
\end

\documentstyle [12pt] {article}
\setlength{\textheight}{10.0in}
\setlength{\textwidth}{7.5in}
\setlength{\oddsidemargin}{-0.5in}
\setlength{\headsep}{0.0in}
\setlength{\topmargin}{-0.75in}
\begin{document}
\pagestyle{empty}
\begin{table}
\begin{center}
\parbox{5.75in}
{\centerline{{\bf TABLE 1}}}
\vskip .1in
\begin{tabular}{|c|c|}
\hline
Reactions&$\langle \Gamma\rangle /F_0$\\ \hline
$\nu_\alpha\bar\nu_\alpha\leftrightarrow e^-e^+$&$8x_W^2\pm 4x_W+1$\\
$\nu_\alpha\bar\nu_\alpha\leftrightarrow \nu_\beta\bar\nu_\beta$&1\\
$\nu_\alpha e^-\leftrightarrow \nu_\alpha e^-$&$8x_W^2\pm 6x_W+{3\over 2}$\\
$\nu_\alpha e^+\leftrightarrow \nu_\alpha e^+$&$8x_W^2\pm 2x_W+{1\over 2}$\\
$\nu_\alpha\nu_\alpha\leftrightarrow \nu_\alpha\nu_\alpha$&6\\
$\nu_\alpha\nu_\beta\leftrightarrow \nu_\alpha\nu_\beta$&3\\
$\nu_\alpha\bar\nu_\alpha\leftrightarrow \nu_\alpha\bar\nu_\alpha$&4\\
$\nu_\alpha\bar\nu_\beta\leftrightarrow \nu_\alpha\bar\nu_\beta$&1\\ \hline
\end{tabular}
\parbox{5.75in} {\baselineskip=20pt
Table 1. The weak reaction rates of neutrinos
averaged over a thermal spectrum. The plus signs correspond to $\alpha=e$,
the minus signs correspond to $\alpha=\mu,\tau$; $\alpha\not= \beta$.}
\end{center}
\end{table}
\end{document}

\documentstyle [12pt] {article}
\setlength{\textheight}{10.0in}
\setlength{\textwidth}{7.5in}
\setlength{\oddsidemargin}{-0.5in}
\setlength{\headsep}{0.0in}
\setlength{\topmargin}{-0.75in}
\begin{document}
\pagestyle{empty}
\begin{table}
\begin{center}
\parbox{6.25in}
{\centerline{{\bf TABLE 2}}}
\vskip .1in
\begin{tabular}{|c|c|c|c|}
\hline
Solutions & Spectral distortion & Neutral Current Event
& Seasonal variation$^1$\\
&(Super-K, SNO)&(SNO)& (Borexino)\\ \hline
Small angle MSW $\nu_e\leftrightarrow\nu_\mu$&Yes&Yes&No\\
Large angle MSW $\nu_e\leftrightarrow\nu_\mu$&No&Yes&Possible\\
Vaccum mixing $\nu_e\leftrightarrow\nu_\mu$&Yes&Yes&Yes\\
Small angle MSW $\nu_e\leftrightarrow\nu_s$&Yes&No&No\\
Vaccum mixing $\nu_e\leftrightarrow\nu_s$&Yes&No&Yes\\
Solar model solution&No&No&No\\ \hline
\end{tabular}
\parbox{6.25in} {\baselineskip=20pt
Table 2. Experimental signatures of different solutions
to the solar neutrino problem in Super-Kamiokande and SNO.\\
$^1$Other than the usual $1/R^2$ variation.}
\end{center}
\end{table}
\end{document}